\documentstyle[12pt,aasms4]{article}
\slugcomment{To Appear in {\it Ap.J}, V. 562, November 20, 2001}
\def\ssp{\def\baselinestretch{1.1}\small\normalsize}
\def\tighttable{\def\baselinestretch{1.0}}
\def\IUE{{\it IUE}}
\def\HST{{\it HST}}
\def\FOS{{\it FOS}}
\def\STIS{{\it STIS}}
\def\ROSAT{{\it ROSAT}}

\def\FUSE{{\it FUSE}}
\def\arcsec{\ifmmode '' \else $''$\fi}
\def\arcmin{\ifmmode ' \else $'$\fi}
\def\arcsecpoint{\ifmmode ''\!. \else $''\!.$\fi}
\def\arcminpoint{\ifmmode '\!. \else $'\!.$\fi}
\def\cc{\ifmmode {\rm cm}^{-3} \else cm$^{-3}$\fi}
\def\cl{\ifmmode {\rm cm}^{-2} \else cm$^{-2}$\fi}
\def\micron{\ifmmode \mu{\rm m} \else $\mu$m\fi}
\def\kms{\ifmmode {\rm km\,s}^{-1} \else km\,s$^{-1}$\fi}
\def\Hubble{\ifmmode {\rm km\,s}^{-1}\,{\rm Mpc}^{-1}
	\else km\,s$^{-1}$\,Mpc$^{-1}$\fi}
\def\ergsec{\ifmmode {\rm ergs\;s}^{-1} \else ergs s$^{-1}$\fi}
\def\ergscm{\ifmmode {\rm ergs\,s}^{-1}\,{\rm cm}^{-2}
	  \else ergs\,s$^{-1}$\,cm$^{-2}$\fi}
\def\ergscmA{\ifmmode {\rm ergs\,s}^{-1}\,{\rm cm}^{-2}\,{\rm \AA}^{-1}
	  \else ergs\,s$^{-1}$\,cm$^{-2}$\,\AA$^{-1}$\fi}
\def\ergscmHz{\ifmmode {\rm ergs\,s}^{-1}\,{\rm cm}^{-2}\,{\rm Hz}^{-1}
	  \else ergs\,s$^{-1}$\,cm$^{-2}$\,Hz$^{-1}$\fi}
\def\Msun{\ifmmode M_{\odot} \else $M_{\odot}$\fi}
\def\Lsun{\ifmmode L_{\odot} \else $L_{\odot}$\fi}
\def\qo{\ifmmode q_{0} \else $q_{0}$\fi}
\def\Ho{\ifmmode H_{0} \else $H_{0}$\fi}

\def\gtsim{\raisebox{-.5 ex}{$\;\stackrel{>}{\sim}\;$}}
\newcommand {\lm}{$\lambda$}

\newcommand {\pg}{{PG~1351+64}}

\newcommand {\lya}{Ly$\alpha$}
\newcommand {\lyb}{Ly$\beta$}
\newcommand{\civ}{C~{\sc iv}}
\newcommand{\siv}{Si~{\sc iv}}
\newcommand{\nv}{N~{\sc v}}

\newcommand{\ciii}{C~{\sc iii}}
\newcommand{\ovi}{O~{\sc vi}}
\newcommand{\he}{He~{\sc ii}}
\newcommand{\mg}{Mg~{\sc ii}}
\newcommand{\oiii}{[O~{\sc iii}]}
\newcommand{\ha}{H$\alpha$}
\newcommand{\hb}{H$\beta$}

\lefthead{Zheng et al.}
\righthead{\pg}

\begin{document}
\ssp

\title{Ultraviolet Broad Absorption Features and the Spectral Energy 
Distribution of the QSO PG 1351+64\altaffilmark{1}}

\author{\sc
W. Zheng\altaffilmark{2},
G. A. Kriss\altaffilmark{2,3},
J. X. Wang\altaffilmark{2,4},
M. Brotherton\altaffilmark{5},
W. R. Oegerle\altaffilmark{2,6}
W. P. Blair\altaffilmark{2},
A. F. Davidsen\altaffilmark{2},
R. F. Green\altaffilmark{5},
J. B. Hutchings\altaffilmark{7},
\& M. E. Kaiser\altaffilmark{2}
}

\altaffiltext{1}{
Based on observations made for the Guaranteed Time Team
by the NASA-CNES-CSA \FUSE\ mission, operated by the Johns Hopkins University
under NASA contract NAS5-32985, 
and observations with the NASA/ESA Hubble Space Telescope, obtained at the
Space Telescope Science Institute, which is operated by the Association of 
Universities of Research in Astronomy, Inc., under NASA contract NAS5-26555.}
\altaffiltext{2}{Center for Astrophysical Sciences, Department of Physics and
	Astronomy, The Johns Hopkins University, Baltimore, MD 21218--2686}
\altaffiltext{3}{Space Telescope Science Institute,
	3700 San Martin Drive, Baltimore, MD 21218}
\altaffiltext{4}{Center for Astrophysics, University of Science and
	Technology of China, Hefei, Anhui, 230026, China}
\altaffiltext{5}{Kitt Peak National Observatory,
	National Optical Astronomy Observatories, P.O. Box 26732,
	950 North Cherry Ave., Tucson, AZ, 85726-6732}
\altaffiltext{6}{Laboratory for Astronomy and Solar Physics,
  Code 681, Goddard Space Flight Center, Greenbelt, MD 20771}
\altaffiltext{7}{Herzberg Institute of Astrophysics, National Research Council
of Canada, Victoria, B. C. V8X 4M6, Canada}
\begin{abstract}

We present a moderate-resolution ($\sim20~\kms$) spectrum of the 
mini-broad-absorption-line QSO \pg\ between 915--1180 \AA,
obtained with the Far Ultraviolet Spectroscopic Explorer ({\it FUSE}). 
Additional low-resolution spectra at longer wavelengths were
also obtained with the Hubble Space Telescope ({\it HST}) and ground-based 
telescopes.
Broad absorption is present on the blue wings of \ciii\ \lm 977,
\lyb, \ovi\ \lm\lm 1032,1038, \lya, \nv\ \lm\lm 1238,1242, 
\siv\ \lm\lm 1393,1402, and \civ\ \lm\lm 1548,1450.
The absorption profile can be fitted with five components at velocities of 
$\sim -780$, $-1049$, $-1629$, $-1833$,
and $-3054$ \kms\ with respect to the emission-line redshift of $z = 0.088$. 
All the absorption components cover a large fraction of the continuum source 
as well as the broad-line region.
The \ovi\ emission feature is very weak, and the \ovi/\lya\ flux ratio 
is 0.08, one of the lowest among low-redshift active galaxies and QSOs. 
The UV continuum shows a significant change in slope near 1050 \AA\ in the 
restframe. The steeper continuum shortward of the Lyman limit extrapolates 
well to the observed weak X-ray flux level. The absorbers' properties are 
similar to those of high-redshift broad absorption-line QSOs.
The derived total column density of the UV absorbers is on the order of 
$10^{21}$ \cl, unlikely to produce significant opacity above 1 keV in the 
X-ray. Unless there is a separate, high-ionization X-ray absorber, 
the QSO's weak X-ray flux may be intrinsic. 
The ionization level of the absorbing components is comparable to that 
anticipated in the broad-line region, therefore the absorbers may be 
related to broad-line clouds along the line of sight. 
\end{abstract}

\keywords{galaxies: active --- galaxies: individual
(PG 1351+64) --- ultraviolet: galaxies}

\section{INTRODUCTION}

Approximately 10\% of QSOs % in magnitude-limited samples
at intermediate and high redshift show strong blueshifted absorption from 
ionized gas, which is being ejected, perhaps by radiation pressure from the 
central continuum source. Ejection velocities as large
as $5 \times 10^4$ \kms\ lead to broad absorption troughs and a 
classic broad-absorption-line QSO spectrum (BALQSO, \cite{bal}; \cite{bal2}). 
Most absorption features are present below 1600 \AA\ in the rest frame. A 
difficulty with studying this phenomenon in high-luminosity QSOs is the 
severe overlapping of the broad troughs that even mask the underlying 
continuum.
The UV spectra obtained with \IUE\ and \HST\ open new windows, enabling one 
to study the absorption features at wavelengths well below 1200 \AA\ 
(\cite{tur}; \cite{kirk}). In low-luminosity QSOs, the velocity range is 
smaller, up to $\sim 4 000$ \kms. Indeed, the absorption features in these 
objects may not be considered as "broad" when compared with those in 
high-luminosity counterparts (\cite{bal3}), but they are significantly
broader amd more displaced than the intrinsic absorption features in 
Seyfert-1 galaxies.  
Thus, these mini-BAL objects offer a possible extension of the BALQSO sample
into the low-redshift domain (\cite{mgh}; \cite{tfw}).
The relatively low velocity structure makes separating and identifying 
individual components in the absorption trough much easier. At low redshifts,
there is less confusion from intervening lines.

The nature and location of the absorbing gas is not known. Since the cores
of emission lines are sometimes completely absorbed, the absorbers must 
lie at least somewhat beyond the broad emission-line region. If the absorption
covering fraction with respect to the nuclear continuum is near 100\%, 
the BAL phenomenon would be unique to a special group
of objects, which would be physically 
different from all other ``normal'' active galactic nuclei (AGN). Conversely,
if the covering factor is small, it is possible that most QSOs contain such
outflows. They are then only recognized as BALQSOs if our line of sight to the
central continuum source happens to pass through the outflow (\cite{bal2}; 
\cite{disk}).
However, \cite{becker} found that the distribution of BAL in their 
radio-selected sample is not consistent with a simple unified model in which
BALQSO are seen edge on.

The broad-band energy distribution of BALQSOs has special characteristics.
It has been known that BALQSOs are not powerful radio sources (\cite{stocke}).
\cite{becker2} found that the number of high-ionization BALQSOs drops 
significantly for QSOs with a radio-loudness parameter $R^* > 100$.
\cite{richards} found a small excess of 
narrow absorption toward radio-loud QSOs and toward 
lobe-dominant QSOs. There is also a small
excess of high-velocity \civ\ absorbers in radio-quiet QSOs as compared to
radio-loud quasars. BALQSOs are weak in the X-ray band (\cite{green}; 
\cite{galla}; Brandt, Laor, \& Wills 2000). The average power-law
index $\alpha_{ox}$ ($f_\nu \propto \nu^{-\alpha}$) derived between 
2500 \AA\ and 2 keV is $\gtsim 1.9$, as compared to $1.5$ for other
low-redshift QSOs (\cite{yuan}). Thus the X-ray level in
BALQSOs is about one order of magnitude weaker than average (\cite{green2}).
The nature of such distinctions is poorly understood, and further studies may
reveal special physical conditions that are intrinsically associated with
the BAL phenomenon.

\pg\ (z=0.088) is one of the few mini-BAL QSOs identified at low 
redshifts. Its \IUE\ spectrum (\cite{prab}) displays significant 
broad absorption in \lya\ and \civ. A tentative analysis
of the \HST\ Faint Object Spectrograph (\FOS) spectra identifies at least 
three absorption components in 
the blue wings of \civ, \lya, and other UV lines (\cite{fos}).  
This object exhibits a weak X-ray flux around 1 keV, with 
a power-law index of $\alpha_{ox} = 1.9$ (\cite{tan}).

In this paper we present the far-UV spectrum of \pg\ obtained with the 
Far Ultraviolet Spectroscopic Explorer (\FUSE),  along with the \FOS\ and 
\STIS\
spectra obtained with the \HST, and optical spectra. The moderate \ovi\ 
absorption and 
the lack of other significant broad low-ionization absorption features suggest
 a moderate 
ionization state in the absorber. The \ovi\ emission is exceptionally weak,
consistent with the prediction of photoionization models with a soft UV-X-ray 
continuum shape. 

\section{DATA}

\FUSE\ is an instrument dedicated to high-resolution spectroscopy in the 
far-ultraviolet spectral region, and it was launched on 1999 June 24. 
It features multiple mirrors, Rowland-circle spectrographs 
and two-dimensional detectors. 
The LiF-coated optics yield a wavelength coverage of 
$\sim 979 - 1187$ \AA, and  the SiC-coated optics cover 
$\sim 905 - 1104$ \AA. For a full description of \FUSE, its mission, and its 
in-flight performance, see Moos et al. (2000)\markcite{moos}, and Sahnow et 
al. (2000)\markcite{sahnow}.

The \FUSE\ observations of \pg\  were carried out between 2000 January 18 
and 20, 
for a total exposure time of 70 ks. The observations were made with the  
$30\arcsec \times 30\arcsec$ low-resolution aperture. Because of a slight
mis-alignment between the optical systems, not all detector channels record
signals that are of photometric quality. Some data in this observation 
suffer from unexpected
abnormalities that have not been fully understood during the calibration
process.

We first combined all the raw data to produce four images in
different detector segments. Two spectra were extracted from the 
two-dimensional data on each segment. These eight extracted spectra were dark 
subtracted and corrected for the background 
stray light. Flux and wavelength calibrations were carried out with the 
standard \FUSE\ calibration pipeline, which assumes a constant value of
dark counts. 
To obtain the best estimate of the zero flux level, we used the Galactic
absorption feature C {\sc ii} 1036 \AA, which appears in all four channels, 
as a calibrator: i.e., the very bottom of this absorption feature represents 
a completely absorbed region, therefore a zero flux level.
We estimate that the flux scale is accurate to $\sim$10\%, and that 
wavelengths are accurate to $\sim$15 \kms. 

We have studied all the sections of data from every channel, and only chosen 
those that are free of abnormalities. The data near the 
edge of the detector segments often show an
abnormal rise in flux and therefore cannot be used.
The spectra were binned by 5 pixels, to preserve the full spectral resolution
of $\sim 20 ~\kms$ for this observation. The spectrum shown in Fig. 1 
is produced by binning the data by 20 pixels (0.12 \AA) to show the overall 
appearance of the spectrum. The S/N level 
around 1080 \AA\ is considerably poorer, because of gaps in the wavelength 
coverage from the various detectors.

As part of a joint \FUSE-\HST\ project, two complementary \HST\ \STIS\ 
snapshot
observations were made on 1999 October 28. The exposure was 600 s for the 
observation with the G140L grating, and 300 s with G230L. Due to the  
short exposures and low source flux, the \STIS\ spectra do not have an 
adequate S/N level for fitting line profiles. We therefore retrieved 
archival \FOS\ spectra of \pg, which has a spectral resolution of 
$\lambda/\Delta\lambda \sim 1300$. The \HST\ spectra between 1150 and 3300 
\AA\ of \pg\ were obtained on 1991 September 5. The exposure time
for the G130H, G190H, and G270H spectra were, respectively, 3200, 3000, and
1400 s. The flux levels of the \STIS\ and \FOS\ spectra match each other 
to within 10\%. 

We used the {\it IRAF} task {\sf specfit} 
(\cite{gak94}) to fit the \FUSE\ and \FOS\ data.  The wavelength scale of 
\FOS\ data was slightly shifted with respect to the Galactic absorption 
features in the spectra, and we made a correction for the offset.
%for all the fitting processes.
The \ovi\ emission features (see Fig. 1) were fitted with four
Gaussian profiles: a narrow doublet and a broad doublet. For the \lyb, \lya, 
\nv, \civ\ and \mg\ emission lines, a narrow component 
and a broad one were assumed for each.
Other emission lines were fitted with one Gaussian profile.
We fixed the widths and velocities of the two components of \lyb\ 
to that of \lya, and linked \nv\ with that of \civ\ (see Table 1). 
The absorption lines were treated as Gaussians in optical depth, and they 
were allowed to partially cover the emission components. Note that the 
covering factor is along the line of sight, not the overall value around the 
source as described in section 1. The wavelengths of the \ovi\ doublets were 
linked at the ratio of their laboratory 
values, and their relative optical depths were fixed at a 2:1 ratio. 
We used five (A, B, C, D \& E) components to fit the \ovi\ absorption 
doublets, and fixed the 
widths and covering factors of the other absorption lines to that of \ovi,  
linking the velocities to that of \ovi\ (a small linear shift in wavelength 
is permitted for lines from different detectors). 
The only exception is component E:  its covering factor cannot be well
constrained by \ovi\ as it is too weak.  We derived the covering  
factor for component E from the \civ\ absorption features in the \FOS\ data. 
Component E of \lyb\ is apparently
narrower than its counterparts in other lines, so we set it free to vary.
Because of different data quality, the \FUSE\ data and \FOS\ spectra were 
fitted separated, with only several linked parameters as discussed above.
Galactic Lyman-line absorption was modeled with the best estimated hydrogen
column density and a varying Doppler parameter. 
An extinction correction of $E_{B-V} = 0.05$ was applied using
the estimated Galactic hydrogen column density of $2.5 \times 10^{20}$ \cl\
(\cite{nh}), and the correction curve of \cite{ccm} with $R_V = 3.1$.
The spectrum between 1100 and 1140 \AA\ is plotted in Fig.~2 to show details 
of the \ovi\ features and the goodness of the fit.  

We obtained optical spectroscopy of PG 1351+640 using the GoldCam spectrograph
mounted on the 2.1-m telescope at Kitt Peak National Observatory.
GoldCam currently uses a Ford 1000 $\times$ 3000 pixel CCD detector, with a
useful area of 400 $\times$ $\sim$2400 pixels in which a long slit spectrum
is focused and can be well calibrated.  
The blue spectrum ($\sim$3200-5900 \AA\ useful coverage) was obtained through 
a wide slit (6$\arcsec$) and using a 500 l\ mm$^{-1}$ grating blazed at 
5500 \AA\ (resolution $\sim$9 \AA) for 300 s on 2000, February 25 (UT).
Weather conditions were generally good on that night, but variable seeing 
(up to 3$\arcsec$) and some thin clouds resulted in non-photometric spectra.
The red spectrum ($\sim$5800-9800 \AA\ useful coverage) was formed by 
combining narrow slit (1\arcsecpoint 5) observations from two nights, both 
taken using a 400 l\ mm$^{-1}$ grating blazed at 8000 \AA\ (resolution 
$\sim$12 \AA) and an OG 550 order-blocking filter: the exposure time on 
February 25 was
400 s, and on February 26 900 s.  The spectra were weighted by their
counts in the combination.  Standard stars were used to divide out 
atmospheric absorption, but the air masses were not well matched and the 
division is imperfect.  The red spectrum flux level was scaled to match that 
of the blue, which in turn was scaled up by a factor of 1.67 to match that 
from the \HST\ spectrum (although the signal-to-noise ratio in the region of
overlap was only 3, making the absolute scaling factor uncertain).
We employed standard data reduction techniques within the NOAO IRAF package.

Archival \ROSAT\ Position-Sensitive Proportional Counter (PSPC) data were 
retrieved from the High Energy Astrophysics Science Archive Research Center 
(HEASARC) at NASA Goddard Space Flight Center. 
We use the data taken in 1993 October, with an exposure time of 3776 s.
The \ROSAT\ spectrum was fitted using {\it xspec} (\cite{xspec})
with a single power law and Galactic absorption. 
The fits generated satisfactory results ($\chi^2_r < 0.8$) for most spectra. 
Our results yield a photon power index $\Gamma = 2.69 \pm 0.2$ 
for a fixed column density of $2.6 \times 10^{20}$ \cl, in good agreement 
with Rush \& Malkan (1996). 
Dual power laws produced only minimal improvements, and therefore
they were not used in the final results. 
Fig. 4 shows the unfolded spectrum with error bars that 
reflect the combined contributions from propagation errors, known flux 
variations, and the uncertainties of the Galactic column density ($10^{19}$ 
\cl). 

\section{Discussion}

\subsection{Emission Lines}

Fig. 5 displays the profiles of seven major UV emission lines and the 
\ha\ line. The fitted parameters for the UV lines are listed in Table 1. 
We will focus our discussion on the features longward of 1050 \AA\ because of
their higher signal-to noise ratios.
 
The most significant finding is the weak \ovi\ emission: its intensity 
measured on the red wing is only about 28\% of that of 
\civ\ or 8\% of that of \lya, which is among the lowest in all AGN and 
QSOs. The \ovi/\lya\ ratio in the QSO composite spectrum of Zheng et al. 
(1997) is only $0.19 \pm 0.02$. Fig. 2 shows the full profile of the \ovi\ 
emission. The broad \ovi\ component appears to be even weaker: 
the \ovi/\lya\ ratio between the broad components is only $\sim 0.02$. 

There is an excess of flux on the blue wings of the Balmer lines, as shown
in Fig. 5.  Such a bump is also  seen 
in another published spectrum (\cite{corbin}). We fitted each of the Balmer 
lines with Gaussian profiles, which are a poor match.
 The fitted profiles are displayed with 
dashed curves. The excess blue flux may be due to  electron scattering or
from the outflowing matter that is part of the broad-line region. 

The \FUSE\ spectra reveal two emission lines at $\sim 1160$ and 1168 \AA, and
we tentatively identify them as S {\sc iv} \lm\lm 1062,1073. Such a feature 
has been noticed, but without a firm identification, by \cite{laor2} (1994, 
1995), and \cite{hamann}. It is also noted as absorption in the spectrum of 
NGC 4151 (\cite{gak4151}). The wavelength of 1168\AA\ coincides 
with the second-order feature of the terrestrial airglow line He {\sc i},
but the feature is much broader than any airglow line. At lower spectral
resolution, such features may have been found as part of the red wing of
the \ovi\ profile in the HST QSO composite spectrum
(\cite{composite}).

The fitted \he\ emission component is very broad, and it includes 
blended features such as an unidentified component at $\sim 1620$\AA\ 
(Laor, et al. 1994, 1995). 
 
\subsection{Absorption Lines}

In Table 2 we list the measurements of major absorption lines, named as 
components A, B, C, D, and E. 
The absorption present in the \FUSE\ data is also obviously present in the
\FOS\ spectrum.
For components A, B, C, D, the line velocities and widths derived from the
\FUSE\ data can also fit the \FOS\ data well, but the derived H {\sc i} 
column densities are quite different (higher than \FOS). This may well be due 
to the lower resolution of \FOS\ data, and saturation in the \lya\ profile.
The presence of \mg\ absorption is highly uncertain as no individual 
components can be seen. 

To determine physical conditions in the absorption components, we used CLOUDY
(Ferland et al. 1998) to calculate models. 
In a grid of photoionization 
calculations, we vary the ionization parameter $U$ from 0.001 to 30. 
We used a  broken power-law ionizing continuum with indices of 
$\alpha=1.9$ and 0.7 between 13.6 eV-2 keV and 2-10 keV, respectively, and 
assumed a 
hydrogen density of $10^{10}$ \cc\ and solar abundances. 
However, variations of these parameters do not affect the results 
significantly.

Because the \FOS\ and \FUSE\  data were not obtained simultaneously, and 
because of the higher spectral resolution
of the \FUSE\ data, we will focus on \FUSE\ data.
For components A, B, C, D, we detected
obvious \ovi\ and \lyb\ absorption. From our grid of photoionization models, 
we determine the
total column density and the ionization parameter $U$ based on the 
observed relative H {\sc i} and \ovi\ column densities. 
With no other constraints, this method can lead to
double-valued results for the ionization parameter and column density
since the ratio of \ovi\ to H {\sc i} will rise to a peak and then decline 
when $U$ increases. 
However, we note that the presence of clear \ciii\ absorption in components 
A, C, D restricts the solutions to the lower ionization parameter in these 
cases, thus yielding small ionization parameters ($U< 0.3$), and a total column
density $< 10^{20}$ \cl. Even if we choose a model with a high 
ionization parameter, the total column density is on the order of $10^{21}$ 
\cl, insufficient to cause significant X-ray obscuration.
Table 3 lists the column densities derived 
ionization parameters for the five absorbers.

\subsection{Continuum}

The continuum in the \FUSE\ data is fitted with a power law of 
$\alpha = 3.92 \pm 0.05$. The fitted power-law index 
for the \FOS\ data is $\alpha = 1.69 \pm 0.01$, which is considerably flatter 
than that for the \FUSE\ data. This suggests
a change in the continuum shape around 1050 \AA\ in the rest frame, 
consistent with but even more dramatic than that found in typical QSOs 
(\cite {composite}; Kriss et al. 2000$a$). To make sure that this is not an 
extinction effect, we carried out a continuum fit to the combined 
\FUSE -\FOS\ spectrum in the wavelength region below 2000 \AA. A simple power 
law with variable extinction yields a significantly poorer fit (best fitted 
value $E_{B-V}=0.06$) than a broken power law with a fixed Galactic 
extinction ($E_{B-V}=0.05$). Therefore, the break in the continuum is 
intrinsic.

Zheng et al. (1997) suggested that 
Compton scattering in a hot corona above the disk 
modifies the exponential turnover of the thermal disk spectrum into a power 
law in the EUV band. In the case of \pg,  a corona with lower temperature 
than average may produce the steep EUV/soft X-ray power law. 
The extrapolation of the soft \FUSE\ continuum leads to the weak
X-ray level that has been observed. \cite{rush} observed a weak
X-ray flux level and a steep X-ray spectrum (photon index $\Gamma \sim
2.6$) in their \ROSAT\ observations pf \pg, which is consistent with an 
extrapolation of our \FUSE\ spectrum. 
The continuum between the UV and soft X-ray bands
can be approximated with a power law with $\alpha_{ox} = 1.9$ (\cite{tan}). 
The short baseline in the sub-\lya\ region does not allow us to assess 
whether the far-UV and the soft X-ray continuum can be approximated with 
a single power law. However, to a first approximation,
such a soft EUV continuum may be 
common among low-luminosity AGN with broad absorption. 
Another similar QSO, PG1411+44, shows an even softer X-ray to UV flux ratio 
with $\alpha_{ox} = 2.1$ (\cite{laor}).
Green \& Mathur (1996) suggested that a torus
of total column density of $N >> 10^{22}$ \cl\ may produce a significant
depression of the soft X-ray flux. But in \pg, the weak soft-X-ray
flux is consistent with the far-UV continuum shape. Furthermore, we do not see
signs for obscuring material of such high column density in the UV or
in the optical data.

Green et al. (1995) analyzed the data of all BALQSOs in the
ROSAT All-Sky Survey and found that they are 
either highly absorbed or underluminous in the soft X-ray bandpass. They 
suggest that the weak X-ray level may be due to heavy obscuration.
\cite{blw} find a significant correlation between the X-ray weakness
in AGN and the strength of the \civ\ absorption and suggest that absorption
may be the cause of the weak X-ray flux. 
In the case of \pg, the photoionization calculations and the lack of an 
intrinsic Lyman edge suggest that the total nucleon column density is less than $10^{22}$ \cl. 
This is not sufficient to obscure the X-ray continuum,
and the soft UV-X-ray continuum is probably intrinsic. 

\subsection{Comparison with Other AGN and QSOs}

It has been suggested that radio-quiet QSOs with weak \oiii\ and strong 
Fe {\sc ii} emission spectra form a class of QSOs that has a high probability
of exhibiting BAL in their spectra (\cite{boroson}; \cite{oiii}). 
In the two BALQSOs 0759+65 (\cite{oiii}) 
and 1700+51 (\cite{tfw}), significant \mg\ absorption and continuum extinction
are also present. In this regard, \pg\ does not fit this 
category: its \oiii\ intensity is about the same as the \hb\ line, the
intensity of its optical Fe {\sc ii} emission is only moderately weak, and no
\mg\ absorption is obvious in \pg\ (Fig. 5).

In terms of UV spectral features, \pg\ bears a strong resemblance to 
PG1411+44.
Their nearly identical redshifts make the two spectra remarkably alike. 
There are significant spectral differences however: 
(1) PG 1411+44 has weaker O [{\sc iii}] emission, by
a factor of about four, with respect to the \hb; and strong optical
Fe {\sc ii} features are present; and (3) PG 1411+44's soft-X-ray flux level 
is significantly lower (by a factor of three or more); and (4) its \he\ \lm 
1640 emission is exceptionally strong, suggesting a harder ionizing 
continuum in the EUV band. 
Laor et al. (1997) found that the X-ray spectrum of PG1411+44 above
2 keV exhibits a significant upturn, suggesting strong 
obscuration in the soft X-ray band.  
Since \pg\ does not present such characteristics, it may belong to a different
group despite the similarity of its \lya\ and \civ\ absorption to that of
PG1411+44.

The  ionization structure of the absorbers in \pg\ seems to be comparable 
to the UV absorption components seen in many Seyfert-1s
(e.g., NGC 4151: 
Kriss et al. 1995; NGC 3516: Kriss et al. 1996$a,b$; Mrk 509: Kriss et al. 
2000$b$) and with those observed in many 
high-redshift objects (\cite{bal}; \cite{kirk}), albeit at a smaller ejection 
velocity. Most of these objects do not show broad \mg\ absorption lines, and
the \civ, \nv, and \ovi\ column densities are at a comparable level to
\pg. More work is needed, however, to understand the difference in their X-ray
properties.

\section{Summary}

\pg\ is one of the handful of mini-BALQSOs.
The \FUSE\ spectrum allows us to identify and measure five kinematically 
distinct components. Model calculations with multiple species suggest column 
densities on the order of a few $ \times 10^{21}$ \cl\ for the absorbing 
clouds. These clouds would not produce
significant absorption in the soft X-ray band. Therefore the object's weak
X-ray flux may be intrinsic unless there is significant absorption from more 
highly ionized material. Additional evidence in support of an intrinsically 
weak X-ray flux exists from the very weak
\ovi\ emission, as well as the significant downturn of the far-UV continuum
shortward of 1050 \AA.

\acknowledgments

This work is based on data obtained for the Guaranteed Time Team by the
NASA-CNES-CSA \FUSE\ mission operated by the Johns Hopkins University.
Financial support to U.S. participants has been provided by NASA contract
NAS5-32985. Additional support for this work is provided in 
part by NASA through Long Term Space Astrophysics grant NAGW-4443 and 
through grant AR-7977.01-96A,
GO 8144 from the Space Telescope Science Institute, which is operated by the 
Association of Universities of Research in Astronomy, Inc., 
under NASA contract NAS5-26555. We thank the referee, Dr. Fred Hamann, for
constructive comments.

Arthur F. Davidsen was a pioneer in ultraviolet astrophysics at Johns Hopkins
University, especially in far-UV observations of extragalactic objects.
He passed away on 2001 July 19, the same day that this paper was accepted
for publication. Among his many prominent roles, he was the Principal 
Investigator for the Hopkins Ultraviolet Telescope, which obtained the first
far-UV spectrum of PG1351+64 in 1995 March during the Astro-2 mission aboard
the space shuttle.
\clearpage
{
\begin{deluxetable}{lccccl}
\tablecaption{UV Emission Lines in \pg. \label{tbl-1}}
\tablewidth{0pt}
\footnotesize
\tighttable
\tablehead{
\colhead{~~~ Line} & \colhead{Wavelength}&
\colhead{Flux\tablenotemark{a}}   & \colhead{FWHM\tablenotemark{b}}&
\colhead{Velocity\tablenotemark{b,c}}& \colhead{Comment}
\\& \colhead{(\AA)}
& \colhead{($10^{-14}$ \ergscm)} & \colhead{(\kms)}
& \colhead{(\kms)}&
}
\startdata
C~{\sc iii} & 977 & $11.9 \pm 2.9$ &$ 2538 \pm 1417$ & $344 \pm 552$\nl % 1064.1 # 46
\lyb\ narrow & 1026 & $5.1\pm 0.4$ &$ 1367$ & $225$ \nl % 1116.6  # 7
\lyb\ broad && $9.4 \pm 0.1$ &$ 7764$ & $-1412 $ \nl % 1116.7 # 6
\ovi\ narrow & 1034& $25.6 \pm 0.5$&$ 2660 \pm 83 $ & $-444 \pm 143$ \nl % 1123.3 # 3
\ovi\ broad&&$6.4 \pm 1.0$ &$ 8597 \pm 995$ & $-407 \pm 114$ \nl % 1123.5 # 4
S {\sc iv}& 1062 & $5.5 \pm 0.2$ &$ 1416 \pm 33$ & $ 617\pm 26$ &\nl % 1157.8  # 33
S {\sc iv}& 1073& $6.7 \pm 0.6$ &$ 1368 \pm 103$ & $ 606 \pm 112$&airglow?\nl % 1168.7  # 34
\lya\ narrow& 1216 & $123 \pm 2$ &$ 1367 \pm 21$ & $225 \pm 10$ \nl % 1323.2 # 5
\lya\ broad &&$330 \pm 5$ &$ 7764 \pm 201$ & $-1412\pm 41$ \nl % 1317.4 # 4
N {\sc v} narrow& 1240 &$17 \pm 2$ &$ 2491\pm 16$ & $129\pm 23$ \nl % 1349.9 # 7
N {\sc v} broad &&$116 \pm 4$ &$ 8198\pm 222$ & $795\pm 49$ \nl % 1343.9 # 6
C {\sc ii} & 1335 & $3.2 \pm 0.4$ &$ 1045 \pm 183$ & $200 \pm 61$ \nl % 1452.9
Si~{\sc iv }+ O~{\sc iv}] &1400& $62 \pm 2$ &$ 4486 \pm 390$ & $-762 \pm 64$ \nl % 1519.3 # 34
\civ\ narrow& 1549 & $57 \pm 1$ &$ 2491 \pm 16$ & $129 \pm 23$ \nl % 1685.8  # 3
\civ\ broad && $74 \pm 1$ &$ 8198 \pm 222$ & $795 \pm 49$ \nl % 1688.9  # 2
% need a manual correction
\he & 1640 & $23 \pm 8 $&$18757\pm 3709$& $4420 \pm $ 1428 & blend\nl 
O~{\sc iii}] &1664&$11\pm 2$ &$ 9312 \pm 1299$ & $-1813 \pm 325$ &\nl 
%N~{\sc iii}] & 1750& $5.6 \pm 0.5$ & $1630\pm 157$&$-1069 \pm 75$ &\nl 
Al~{\sc iii}&1857&$12 \pm 1$ &$ 4021 \pm 1123$ & $1010 \pm 607$ &\nl 
Si~{\sc iii}] & 1892& $17 \pm 2$ &$ 3406 \pm 995$ & $ 358 \pm 393$ &\nl 
C~{\sc iii}] & 1909& $29 \pm 3$ &$ 2853 \pm 724$ & $ 551 \pm 321$ &\nl
Mg~{\sc ii} narrow & 2798& $7 \pm 4$ &$ 1367 \pm 207$ & $195 \pm 120$ &\nl 
Mg~{\sc ii} broad& & $57 \pm 8$ &$ 7407 \pm 591$ & $3\pm 137$& Fe {\sc ii}?\nl 
\enddata
\tablenotetext{a}{Corrected for $E_{B-V}=0.05$.} 
\tablenotetext{b}{Values without errors are fixed.}
\tablenotetext{c}{With respect to the systemic velocity of 26,400~\kms\
(Malkan et al. 1987).}
\end{deluxetable}
}

\clearpage
\begin{deluxetable}{lcccccc}
\tablecaption{UV Absorption Lines in \pg} %\label{tbl-2}}
\tablewidth{0pt}
\tablehead{
\colhead{~~~ Line} & \colhead{Wavelength} 
& \colhead{Component}&
% \colhead{Optical depth}   
\colhead{Column density} 
& \colhead{FWHM}    &
\colhead{Velocity\tablenotemark{a}}&
\colhead{Covering factor\tablenotemark{b}}\\
& \colhead{(\AA)} & 
& \colhead{($10^{14}$\cl)}  
%&&\colhead{at line center\tablenotemark{a,b}} 
& \colhead{(\kms)}
& \colhead{(\kms)} &
}
\footnotesize
\startdata
C{\sc iii}&977.02&A&$1.81\pm 0.64$ &$198\pm 15$&$-780\pm 23$&$0.71\pm 0.02$\nl % 1060.3  # 47
&&B&$0.44\pm 0.28$&$250\pm 10$&$-1049\pm 8$ & $0.60 \pm 0.04$ \nl % 1059.3  # 48
&&C&$0.37\pm 0.12$ &$ 135\pm 8$ & $-1629\pm 10$ & $0.80 \pm 0.04$ \nl % 1057.3  # 49
&&D&$1.33\pm 0.62$ &$ 134\pm 16$ & $-1833\pm 8$ & $0.75 \pm 0.02$ \nl % 1056.6  # 55
&&E&$2.99\pm 0.56$ &$ 613\pm 60$ & $-3054\pm 125$ & $0.87$ \nl % 1054.2  # 50
\lyb &1025.72 &A& $11.7 \pm 1.9$ &$ 198\pm 15$&$-780\pm 23$ & $0.71 \pm 0.02$ \nl % 1113.1  # 23
&&B& $6.14\pm 1.90$ &$ 250\pm 10$ & $-1049\pm 8$&$0.60 \pm 0.04$ \nl % 1112.1  # 25
&&C & $4.29 \pm 2.54$ &$ 135\pm 8$ & $-1629\pm 10$&$0.80 \pm 0.04$ \nl % 1110.0  # 26
&&D & $0.93 \pm 0.36$ &$ 134\pm 16$ & $-1833\pm 8$ & $0.75 \pm 0.02$ \nl % 1109.2  # 27
&&E & $1.89 \pm 0.42$ &$ 65\pm 8$ & $-3054\pm 125$ & $0.87$ \nl % 1106.7  # 28
\ovi &1031.93,1037.62 & A&$26.2\pm 2.8$ &$198 \pm 15$ & $-780 \pm 23$ & $0.71 \pm 0.02$ \nl % 1126.0 #9,16
 & &B&$41.5\pm 8.7$ &$250\pm 10$ & $-1049\pm 8$ & $0.60\pm 0.04$ \nl % 1125.0 #11
 & &C &  $7.06\pm 2.02$ &$ 135 \pm 8$ & $-1629 \pm 10$ & $0.80\pm 0.04$ \nl % 1122.8  # 12
 & &D & $13.9\pm 8.9$ &$ 134\pm 16$ & $-1833\pm 8$ & $0.75\pm 0.02$ \nl % 1122.1  # 13
&&E& $0.96\pm 0.59$&$613 \pm 60$ & $-3054\pm 125$ & $0.87$ \nl % 1119.6  # 14
\lya&1215.67&A&$3.32 \pm 0.17$ &$ 407\pm 14$ & $-779\pm 17$ &$0.71$ \nl % 1318.8  # 22
 &  &B&$0.48\pm 0.20$ &$189\pm 25$ & $-1479 \pm 10$ & $0.60$ \nl % 1315.8  # 20
&&C& $0.11 \pm 0.08$ &$ 114\pm 8$ & $-1633 \pm 7$ & $0.80$ \nl % 1314.9  # 19
 & &D &$0.72 \pm 0.33$ &$ 234\pm 14$ & $-1964 \pm 11$ & $0.75$ \nl % 1314.6  # 21
 & &E & $1.26 \pm 0.15$ &$ 587\pm 17$ & $-3132\pm 10$ & $0.87 \pm 0.02$ \nl % 1309.6  # 18
N {\sc v}&1238.81,1242.80&A&$8.90\pm 0.41$ &$ 497\pm 14$ & $-869\pm 17 $ &$0.71$\nl%1343.9#27
& &B&$1.06 \pm 0.25$ &$ 189\pm 25$ & $-1479 \pm 10$ & $0.60$ \nl % 1339.6  # 26
&&C& $1.71 \pm 0.36$ &$ 114\pm 8$ & $-1633 \pm 7$ & $0.80$ \nl % 1340.8  # 25
& & D&$1.14 \pm 0.24$ &$ 234\pm 14$ & $-1964 \pm 11$ & $0.75$ \nl % 1340.1  # 24
& &E&$6.17 \pm 0.69$ &$ 587\pm 17$ & $-3132\pm 10$ & $0.87 \pm 0.02$ \nl % 1334.6  # 23
\siv&1393.76,1402.77&A&$0.80\pm 0.24$&$497\pm 14$&$-869\pm 17$&$0.71$\nl%1512.0#39
&&B&$0.19\pm 0.23$&$189\pm 25$ & $-1479 \pm 10$ & $0.60$ \nl % 1511.2 # 40?
&&C&$0.08\pm 0.06$&$114\pm 8$&$-1633 \pm 7$& $0.80$\nl %1508.5 #39>10
&&D&$0.25 \pm 0.30$ &$ 234\pm 14$ & $-1964 \pm 11$ & $0.75$ \nl % 1507.1  # 38>9
&&E&$0.39 \pm 0.30$ &$ 587\pm 17$ & $-3132\pm 10 $ & $0.87\pm 0.02$ \nl % 1508.5#37>8
\civ&1548.19,1550.77&A&$5.65\pm 0.16$ &$407\pm 14$&$-869\pm 17$&$0.71$ \nl % 1679.6 #12
 &&B&$1.02 \pm 0.17$ &$ 189 \pm 25$ & $-1569 \pm 10$ & $0.60$ \nl % 1675.7  # 10
 &&C&$1.43 \pm 0.16$ &$ 114 \pm 8$ & $-1723 \pm 7$ & $0.80$ \nl % 1674.8  # 9+11
 &&D&$1.02 \pm 0.17$ &$ 234 \pm 14$ & $-2054 \pm 11$ & $0.75$ \nl % 1670.2  # 13
 &&E&$3.64 \pm 0.81$ &$ 587 \pm 17$ & $-3222\pm 10$ &$0.87\pm 0.02$ \nl % 1667.8 # 8
%%%%%%%%%%%%%%%%%%%%%%%%
%\mg & 2799 & &  $1.9 \pm 1.6$ &$ 2317 \pm 724$& $-3506 \pm 101$ & \nodata \nl 
% &  &A&  $2.4 \pm 0.5$ &$ 2911 \pm 335$& $-2485 \pm $ 456&\nl 
% & &B&  $ 2.2 \pm 1.0$ &$ 2540 \pm 290$& $-2269 \pm 595$ &\nl 
% &  &C&  $2.2 \pm 0.5$ &$ 1898 \pm 417$& $-1135 \pm $ 276 &\nl  
% & & D&$ \pm $ & & $$ &\nl 
%  &  & &  $5.0 \pm 2.2$ &$2626\pm 556$& $ 2081 \pm 179$ &\nl 
%  &  & &  $2.8 \pm 0.2$ &$ 3462 \pm 811$& $3269 \pm 370$ &\nl 
\enddata
%\tablenotetext{a}{Corrected for $E_{B-V}=0.05$}. 
\tablenotetext{a}{With respect to the systemic velocity of 26,400~\kms.}
\tablenotetext{b}{Values without errors are fixed.}
%\tablenotetext{b}{For doublets, values refer to the stronger component.}
%\tablenotetext{d}{Estimated for the entire line.}
%\tablenotetext{e}{Corrected for instrumental profile.}
%\tablenotetext{f}{\FOS\ data, normalized to the HUT data.}
\end{deluxetable}

\clearpage
\begin{deluxetable}{lccccc}
\tablecaption{Properties of the Absorbers in \pg. \label{tbl-3}}
\tablewidth{0pt}
\tablehead{
\colhead{Species}&
\multicolumn{5}{c}{Component} \\
&\colhead{A} & \colhead{B} &\colhead{C}&\colhead{D}&\colhead{E}}
\footnotesize
\startdata
&\multicolumn{5}{c}{Column density(\cl)} \\ \hline
H{\sc i}&$1.1 \times 10^{15}$&$6.1\times 10^{14} $&$4.2\times 10^{14}$&$8.2\times 10^{13} $& $1.3\times 10^{14}$ \\
C{\sc iii}&$1.8\times 10^{14}$&$4.3\times 10^{13}$ &$3.7 \times 10^{14}$&$1.3\times 10^{14}$&$3.0\times 10^{14}$  \\
%C{\sc iv}&$2.3\times 10^{19} $&$3.6\times 10^{18}$&$1.9\times 10^{19}$&$2.3\times 10^{18} $&$4.4\times 10^{18}$  \\
O{\sc vi}&$2.62 \times 10^{15}$&$4.2\times 10^{15}$&$7.1\times 10^{14}$&$1.4\times 10^{15}$&$1.0\times 10^{14}$ \\ \hline
%N{\sc v}&$5.7\times 10^{19} $&$3.9\times 10^{18}$&$2.7\times 10^{19}$&$3.1\times 10^{18} $&$1.3\times 10^{19}$  \\
%Si{\sc iv}&$6.7\times 10^{18}$&$1.1\times 10^{18}$&$1.8\times 10^{18} $& $2.6\times 10^{16}$& $1.1\times 10^{17}$ \\ \hline
$\log U$&$-1.19$ &$-1.00$ &$-1.23$ &$-0.74$ &$-1.23$ \\ \hline
$N_{total}$ (\cl)&$2.1 \times 10^{19}$ &$1.8\times 10^{19}$ & $7.3\times 10^{18}$&$4.4\times 10^{18}$ &$9.6\times 10^{17}$ \\\hline
Covering factor& 0.71& 0.60&0.80 & 0.75& 0.87\\ \hline
\enddata
\end{deluxetable}

\clearpage

\centerline{\bf Figure Captions}
\bigskip

\figcaption{{\it FUSE} spectrum of \pg, binned to $\sim 0.12$ \AA\ (20 pixels).
$1\sigma$ errors are displayed as dotted lines. 
%Significant residual features due to airglow lines are marked with the 
%Earth symbols. 
The data between
1070 and 1100 \AA\ are not as reliable because they lie in a gap of several
detector segments.
% The wavelength for the assumed Lyman limit edge is marked. 
\label{fig1}}
\bigskip

\figcaption{Best fit to the O {\sc vi} emission and absorption features.
The data are binned by five pixels. The residuals are displayed at the bottom.
% Possible interstellar absorption features are marked with an ``@'' sign. 
The centroid wavelengths of the fitted O~{\sc vi} and Ly$\beta$ absorption 
components are marked with their assigned names. 
\label{fig2}}
\bigskip

\figcaption{Combined {\it STIS} and {\it FUSE} Spectrum of PG1351+64.  
The {\it STIS} data below 1750 \AA\ are not used due to abnormalities, and 
the {\it FOS} G190H spectrum
is overplotted to fit the wavelength gap. Galactic absorption features are
marked with a "G".
\label{fig3}}
\bigskip
\figcaption{ 
Broad-band energy distribution of \pg. Corrections are
made for a hydrogen column density of $2.5 \times 10^{20}$ \cl\ and extinction
of $E_{B-V} = 0.05$. 
\label{fig4}}
\bigskip

\figcaption{Major absorption features and their absorption counterparts. 
The vertical dashed lines mark the position of five absorption components, 
and   the arrow signs in panel $a,c,f$ and $g$ represent the other doublet 
components.
The zero level approximately represents that of the fitted continuum.
\label{fig5}}

\clearpage\setcounter{figure}{0}
\begin{figure}
\plotfiddle{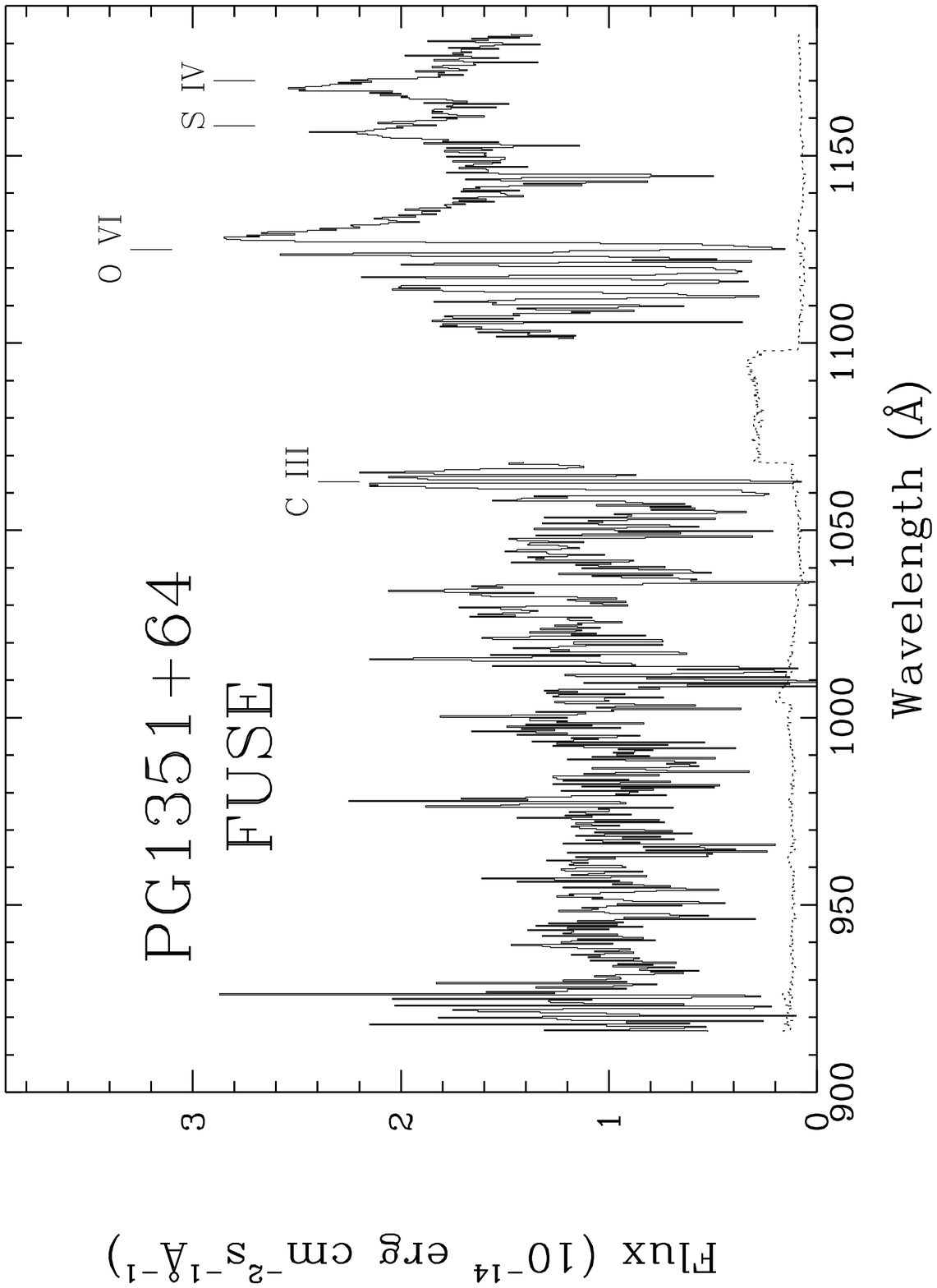}{7in}{0}{85}{85}{-260}{-30}
\caption{~}
\end{figure}

\begin{figure}
\plotfiddle{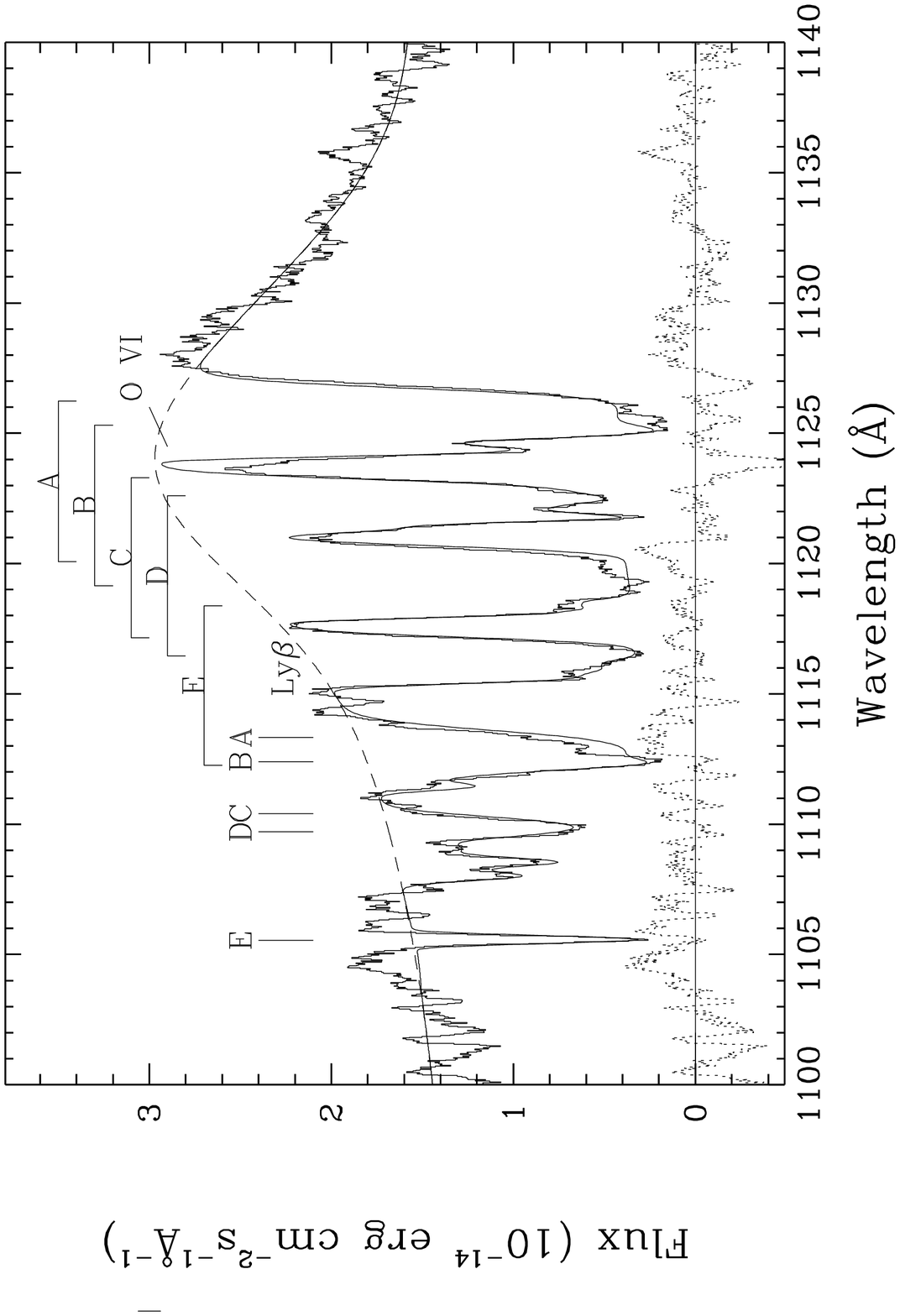}{7in}{-90}{75}{75}{-275}{500}
\caption{~}
\end{figure}

\begin{figure}
\plotfiddle{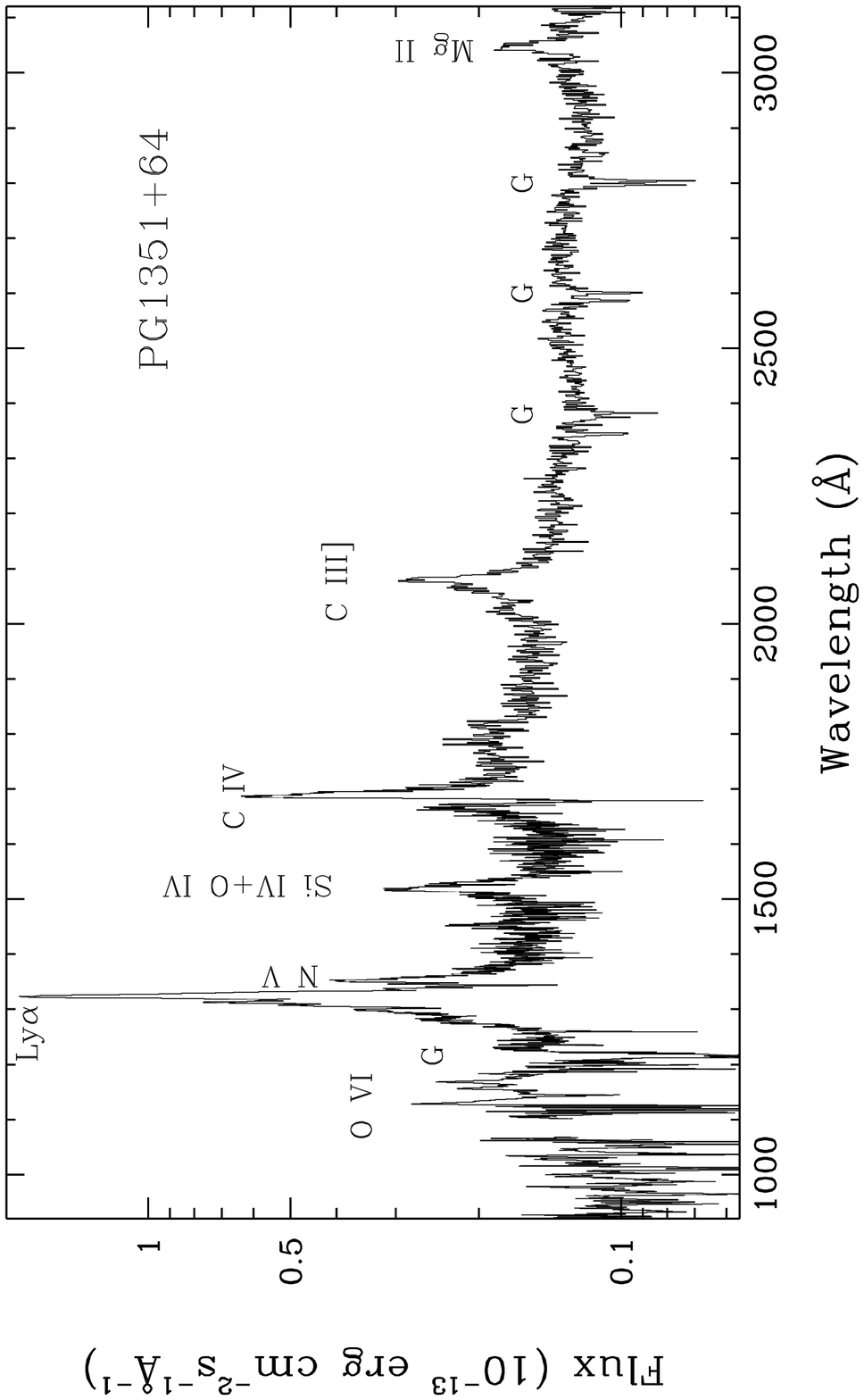}{7in}{0}{85}{85}{-230}{-60}
\caption{~}
\end{figure}

\begin{figure}
\plotfiddle{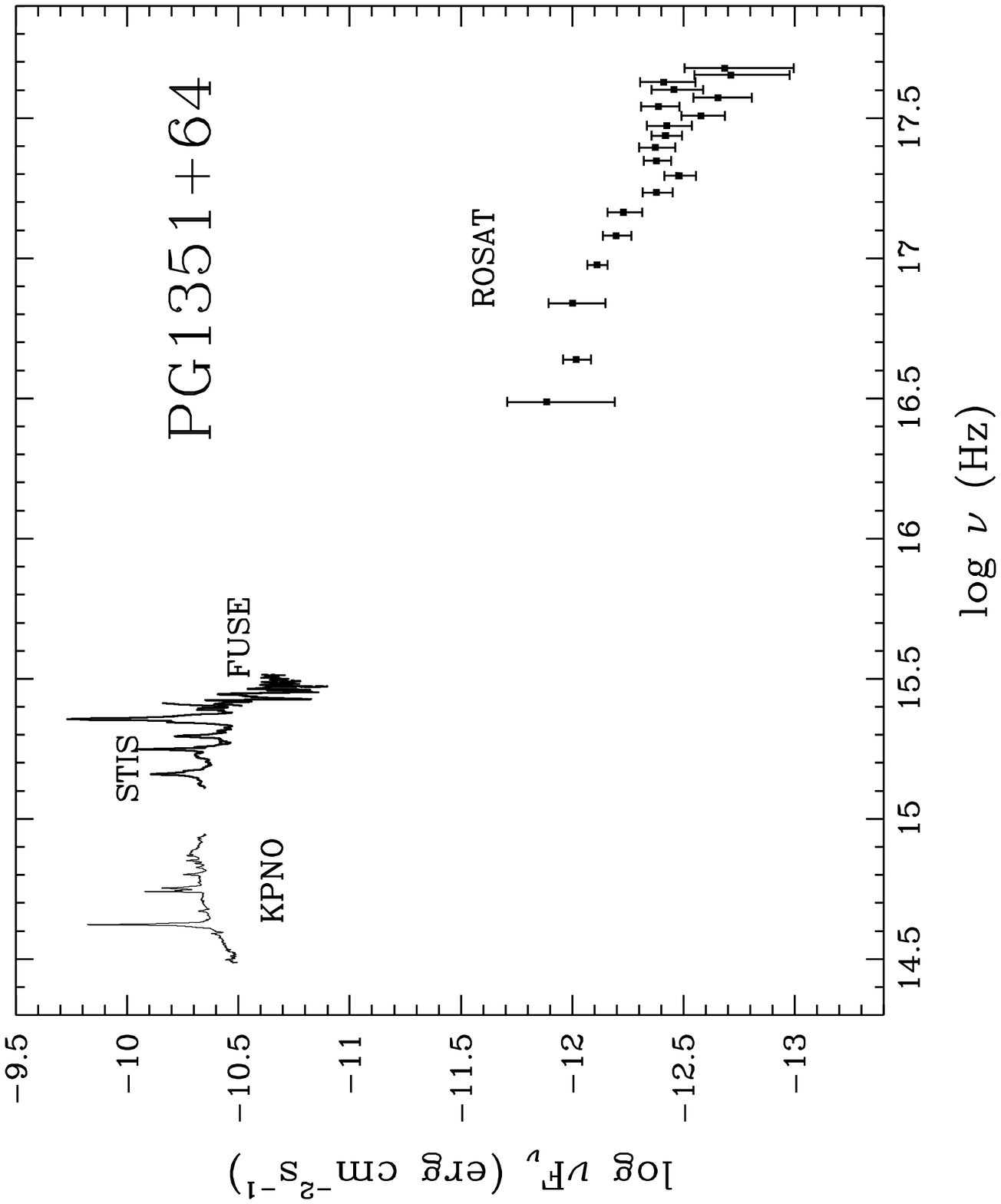}{7in}{0}{85}{85}{-260}{-30}
\caption{~}
\end{figure}

\begin{figure}
\plotfiddle{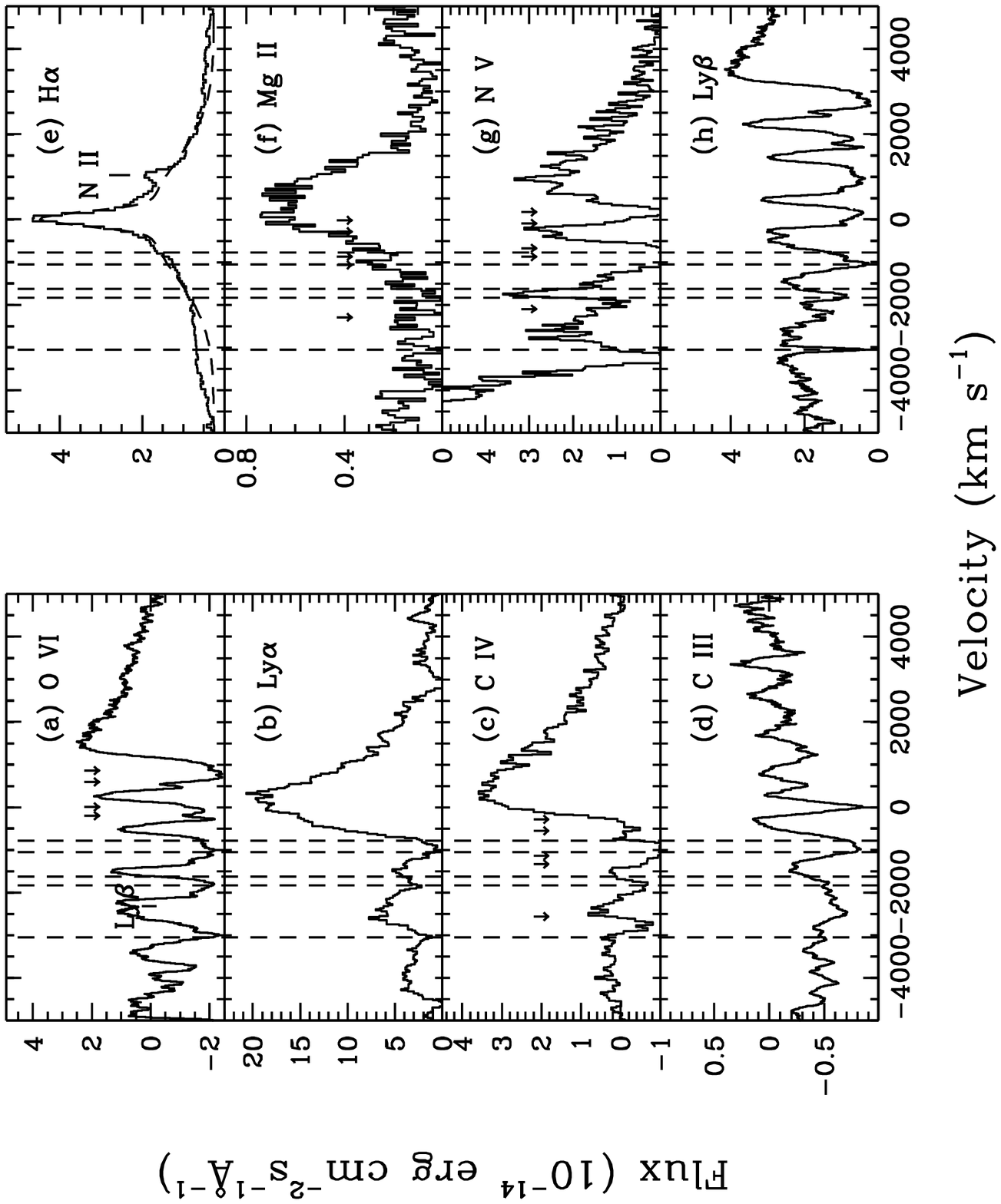}{7in}{0}{85}{85}{-260}{-30}
\caption{~}
\end{figure}

\end{document}